\documentclass[twocolumn,PRB,aps,psfig,showpacs,preprintnumbers,superscriptaddress]{revtex4-2}
\usepackage{graphicx}
\usepackage{epstopdf}
\usepackage{float}
\usepackage{color}
\usepackage{amsmath}
\usepackage{bm}

\begin{document}
\title {Role of Nematic Fluctuations on Superconductivity in FeSe$_{0.47}$Te$_{0.53}$ Revealed by  NMR under Pressure }

\author{Qing-Ping Ding}
\affiliation{Ames National Laboratory, U.S.DOE, Ames, Iowa 50011, USA}
\author{Juan Schmidt}
\affiliation{Ames National Laboratory, U.S.DOE, Ames, Iowa 50011, USA}
\affiliation{Department of Physics and Astronomy, Iowa State University, Ames, Iowa 50011, USA}
\author{ Jose A. Moreno}
\affiliation{Ames National Laboratory, U.S.DOE, Ames, Iowa 50011, USA}
  \affiliation{Departamento de Física de la Materia Condensada, Instituto Nicolás Cabrera and Condensed Matter Physics Center, Universidad Autónoma de Madrid, E-28049 Madrid, Spain}
\author{Sergey L. Bud'ko}
\affiliation{Ames National Laboratory, U.S.DOE, Ames, Iowa 50011, USA}
\affiliation{Department of Physics and Astronomy, Iowa State University, Ames, Iowa 50011, USA}
\author{Paul C. Canfield}
\affiliation{Ames National Laboratory, U.S.DOE, Ames, Iowa 50011, USA}
\affiliation{Department of Physics and Astronomy, Iowa State University, Ames, Iowa 50011, USA}
\author{Yuji Furukawa}
\affiliation{Ames National Laboratory, U.S.DOE, Ames, Iowa 50011, USA}
\affiliation{Department of Physics and Astronomy, Iowa State University, Ames, Iowa 50011, USA}

\date{\today}

\begin{abstract} 
The relationship between antiferromagnetic (AFM) spin fluctuations (SF), nematic fluctuations, and superconductivity (SC) has been central to understanding the pairing mechanism in iron-based superconductors (IBSCs).
     Iron chalcogenides, which hold the simplest crystal structure in IBSCs, provide a good platform to investigate the relationship.
     Here, we report $^{77}$Se and $^{125}$Te nuclear magnetic resonance studies of FeSe$_{0.47}$Te$_{0.53}$, which is located close to a  nematic quantum critical point (QCP), under pressures up to 1.35 GPa.
    Both the superconducting critical temperature and AFMSF were found to be enhanced under pressure, which suggests a correlation between SC and AFMSF in FeSe$_{0.47}$Te$_{0.53}$. 
    However,  the contribution of AFMSF to SC in FeSe$_{0.47}$Te$_{0.53}$ was found to be much less compared to that in FeSe$_{1-x}$S$_{x}$, suggesting that  nematic fluctuations play a dominant role in the SC in FeSe$_{1-x}$Te$_{x}$ around the nematic QCP.

\end{abstract}

\maketitle

    The interplay between magnetic fluctuations, electronic nematicity and the unconventional superconductivity (SC) has received wide interest after the discovery of SC in iron pnictides \cite{Kamihara2008}.
    In many of the iron pnictide superconductors, by lowering temperature  ($T$), the crystal structure changes from high-$T$ tetragonal ($C_4$ symmetry)  to low-$T$ orthorhombic ($C_2$ symmetry) around a system-dependent N\'eel temperature $T_{\rm N}$, below which long-range stripe-type antiferromagnetic (AFM) order emerges \cite{Canfield2010,Johnston2010,Stewart2011}.
Nematicity is associated with this structural transition that breaks $C_4$ symmetry and is characterized by the development of in-plane anisotropy in the electronic properties.
   SC in these compounds emerges upon suppressing both the structural (or nematic) and magnetic transitions by carrier doping and/or applying pressure ($p$).
    Although this suggests a close relationship between AFM and nematic phases, it also makes a difficulty to separate the individual contribution to SC. 
    
  In this context,  CaK(Fe$_{1-x}$$M$$_{x}$)$_4$As$_4$ ($M$ = Co, Ni, Mn, Cr) is a rare and novel SC system which exhibits only AFM state called a hedgehog spin-vortex crystal (HSVC) without nematic phase transition \cite{Cui2017,Meier2018,Wilde2023,Xu2023}.
    From nuclear magnetic resonance (NMR) measurements, a possible HSVC AFM quantum critical point (QCP) has been reported around $x$ $\sim$ 0 where the AFM spin fluctuations (SF) play an important role in the appearance of SC \cite{Ding2018}. 

\begin{figure*}[tb]
\includegraphics[width=\textwidth]{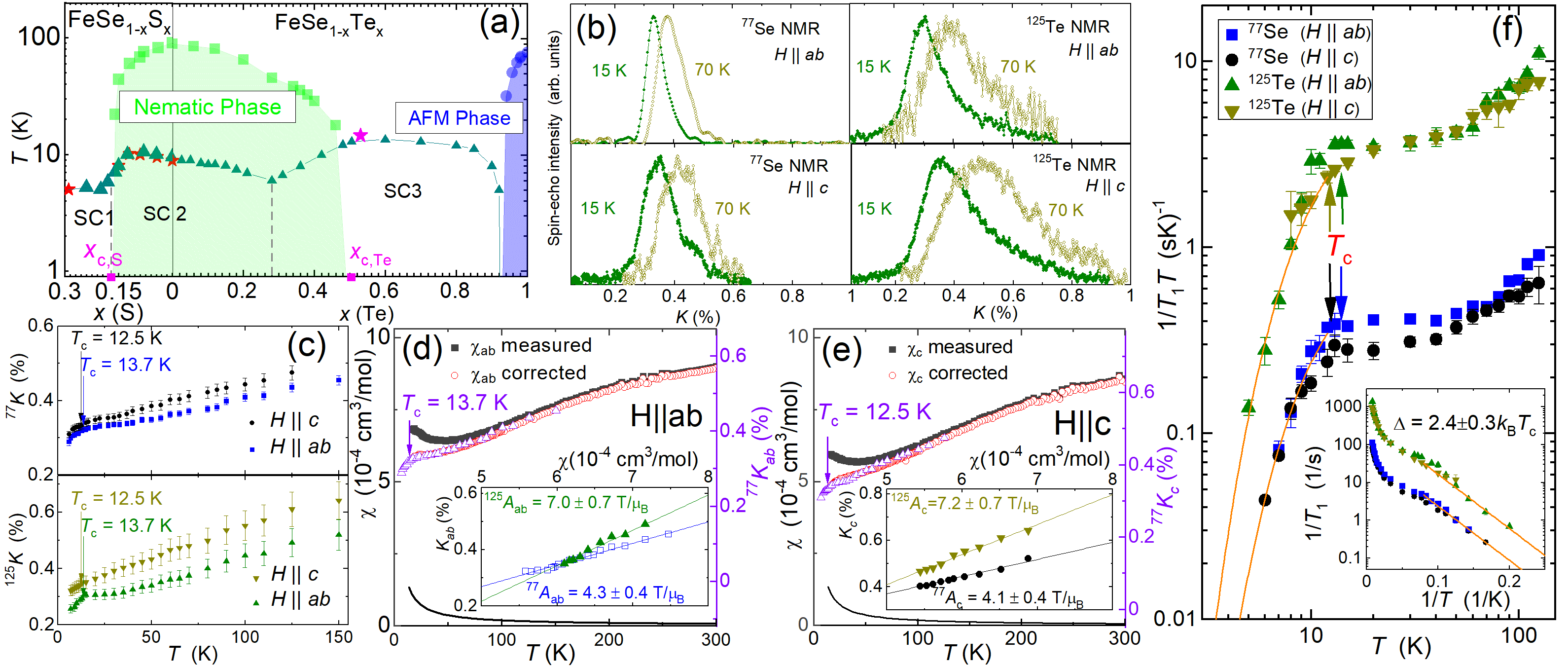} 
\caption{(a) Phase diagrams in FeSe$_{1-x}$Te$_{x}$ and FeSe$_{1-x}$S$_{x}$ (note the vertical axis is in log-scale).  
 $x_c$ indicates the nematic QCP.  
SC1, SC2, and SC3 are three different SC regions \cite{Mukasa2023}. The data in triangles ($T_c$) and squares ($T_{\rm {nem}}$) are from Ref. \cite{Mukasa2021}, the data in circles ($T_{\rm N}$) are from Ref. \cite{Otsuka2019}, and the data in stars($T_{\rm c}$) are from Ref. \cite{Wiecki2018} and the present study.
(b) $^{77}$Se and $^{125}$Te NMR spectra of FeSe$_{0.47}$Te$_{0.53}$ at $T$ = 15 K and 70 K for $H$ $\parallel$ $ab$ and $H$ $\parallel$ $c$. 
(c) $T$ dependencies of the $^{77}$Se and $^{125}$Te Knight shifts $K$.
$\chi (T)$  and $^{77}K$ (right axes) for $H$ $\parallel$ $ab$ and  $H$ $\parallel$ $c$ are shown in (d) and (e), respectively. 
The upturns in $\chi(T)$ below $\sim$ 50 K are due to impurities that follow Curie behavior $\chi_{\rm imp}$ = $C/T$ with $C$ = 0.0018 cm$^3$/mol for both directions, as shown in black lines.
The open circles are corrected $\chi(T)$ by subtracting $\chi_{\rm imp}$ (See text).
Insets: $K(T)$ versus $\chi(T)$ for the corresponding $ab$ and $c$ components of $K$.
The solid lines are linear fits.
(f) $T$ dependencies of $^{77}$Se and $^{125}$Te 1/$T_1T$.  The inset shows the semi-log plot of  $1/T_1$ vs. 1/$T$.
The orange solid curves below $T_{\rm c}$ are calculated results using a full-gap model.
}
\label{fig:Fig1}
\end{figure*}

        The S or Te substituted FeSe system  also provides a favorable platform for the study of the impact of nematicity or antiferromagnetism on SC independently \cite{Bohmer2018,Coldea2019,Mukasa2021,Hosoi2016, Ishida2022}. 
    In contrast to the CaK(Fe$_{1-x}$$M$$_{x}$)$_4$As$_4$ system,  the superconductor FeSe ($x$ = 0) with a critical temperature of $T_{\rm c}$ = 8.5 K exhibits only a nematic phase transition at $T_{\rm nem}$ = 90 K without AFM ordering at ambient $p$ \cite{Hsu2008,McQueen2009,Bohmer2018}.  
     S or Te substitutions for Se in FeSe result in intriguing phase diagrams as shown in Fig.~\ref{fig:Fig1}(a). 
     
     In the case of the S substitution, the nematic phase [the green region in  Fig.~\ref{fig:Fig1}(a)] is gradually suppressed and disappears around  a nematic QCP  $x_{c, {\rm S}}$ $\sim$  0.17 \cite{Hosoi2016}.  
      In contrast, $T_{\rm c}$ first increases from $T_{\rm c}$ = 8.5 K up to 10 K around $x$ = 0.09 \cite{Abdel2015,Watson2015,Reiss2017}, 
then drops to $T_{\rm c}$ $\sim$ 5 K at $x_{c, {\rm S}}$ without any enhancement of $T_{\rm c}$  around the nematic QCP [SC2 region in  Fig.~\ref{fig:Fig1}(a)].  
   Above  $x_{c, {\rm S}}$,  $T_{\rm c}$  becomes nearly independent of $x$ [SC1 region  in Fig.~\ref{fig:Fig1}(a)] and   the fully replaced FeS is still a superconductor with $T_{\rm c}$ = 5 K \cite{Lai2015}.  
       As in the case of FeSe, no AFM state has been observed in FeSe$_{1-x}$S$_{x}$ at ambient $p$.  
      Nevertheless,  NMR measurements reveal clear correlations between $T_{\rm c}$ and nuclear spin-lattice relaxation rate $1/T_1$ \cite{Böhmer2015,Baek2015, Imai2009,Wiecki2017,Wiecki2018}, but in different ways in the $C_2$ and ${C_4}$ phases \cite{Rana2020,Rana2022, Rana2023,Kuwayama2021,Kuwayama2019,Kuwayama2020,Yu2023},  showing the importance of spin fluctuations in FeSe$_{1-x}$S$_{x}$.

     In the Te-substitution case,  $T_{\rm c}$ initially decreases to a local minimum at $x$ $\sim$ 0.3 and then starts increasing towards the board maximum around $x$ = 0.6,  making two different SC regions (SC2 and SC3) as shown in Fig.~\ref{fig:Fig1}(a) in  FeSe$_{1-x}$Te$_{x}$ \cite{Ishida2022}. 
 It is reported that,  from the elastoresistivity and upper critical field measurements on  FeSe$_{1-x}$Te$_{x}$,  
 the enhancement of  $T_{\rm c}$ is around the pure nematic critical point  $x_{c, {\rm Te}}$ $\sim$ 0.5  where the nematic phase disappears and a diverging nematic susceptibility was observed \cite{Mukasa2023,Ishida2022,Jiang2023}. 
    At very close to FeTe above $x$  $\sim$ 0.9,  the AFM state with the bicollinear magnetic structure can be observed where $T_{\rm N}$ increases from 32 K at $x$ = 0.94 to 76 K at $x$ = 1 \cite{Otsuka2019}.
    The two distinct SC regions were clearly demonstrated by applying magnetic field $H$  \cite{Mukasa2023}.
   At $\mu_0H$ = 14 T, $T_{\rm c}$ around $x$ = 0.3 in  FeSe$_{1-x}$Te$_{x}$ is strongly suppressed leading to two split SC regions originating from SC2 and SC3 (note SC1 is completely suppressed at this $H$) \cite{Mukasa2023}.  
   The SC2 regions shrunk leading to a SC dome in S-substituted region centered around $x$ = 0.1 on FeSe$_{1-x}$S$_{x}$ with $T_{\rm c}$ $\sim$ 2 K is observed, and it disappears completely at $\mu_0H$ = 30 T \cite{Mukasa2023}.  
     In contrast,  the SC dome centered at $x$ = 0.6 in  FeSe$_{1-x}$Te$_{x}$ persists at 30 T and, survives even at $\mu_0H$ = 46 T  at $x$ close to the nematic QCP, suggesting that nematic fluctuations play an important role in the appearance of SC in FeSe$_{1-x}$Te$_{x}$ near   $x_{c, {\rm Te}}$ \cite{Mukasa2023,Ishida2022}.
 It is also interesting to point out that,  although FeSe$_{1-x}$Te$_{x}$ has been studied less than FeSe$_{1-x}$S$_{x}$ due to its higher inhomogeneity originating from  the difficulty in synthesizing  FeSe$_{1-x}$Te$_{x}$ single crystals \cite {Hou2023}, 
FeSe$_{1-x}$Te$_{x}$ has been considered as one of the platforms for topological superconductivity, evidenced by scanning tunneling spectroscopy and angle-resolved photoemission spectroscopy measurements \cite{Yin2015,Wang2018,Machida2019,Zhang2018,Li2024}.
  These characteristics also underline the importance of understanding the superconducting mechanism in FeSe$_{1-x}$Te$_{x}$.

  It is therefore crucial to investigate in detail how the magnetic fluctuations correlate with $T_{\rm c}$ in FeSe$_{1-x}$Te$_{x}$.
Usually,  such studies can be performed by NMR measurements on different Te content samples with different $T_{\rm c}$ in FeSe$_{1-x}$Te$_{x}$.  
On the other hand,  it has been shown that $T_{\rm c}$ largely depends on $p$ in FeSe$_{1-x}$Te$_{x}$ \cite{Mukasa2021},  which provides a good opportunity to study it without changing samples.  
In this Letter, we report the results of $^{77}$Se and  $^{125}$Te NMR measurements on FeSe$_{0.47}$Te$_{0.53}$, which is close to the  nematic QCP [Fig.~\ref{fig:Fig1}(a)], to investigate the relationship between SC,  nematic fluctuations, and AFMSF.  
     Owing to the sensitivity of $T_{\rm c}$ on $p$, by carrying out NMR measurements up to 1.35 GPa,  we found that  the contribution of AFMSF to SC in FeSe$_{0.47}$Te$_{0.53}$ is much less compared to that in FeSe$_{1-x}$S$_{x}$, and nematic fluctuations can be the source of SC in  FeSe$_{0.47}$Te$_{0.53}$.

\begin{figure*}[tb]
\includegraphics[width=\textwidth]{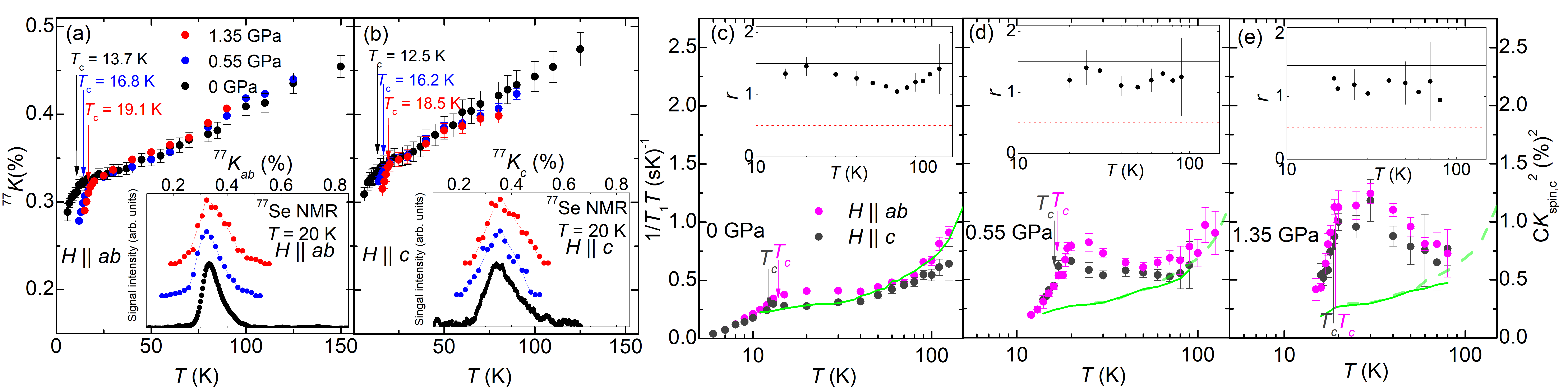} 
\caption{$T$ dependencies of the $^{77}$Se $K_{ab}$  (a) and $K_{c}$ (b) under various $p$. 
Downward arrows indicate $T_{\rm c}$ under $\mu_0H$ = 7.4089 T for each $H$ direction.
Insets show the $^{77}$Se NMR spectra at $T$ = 20 K under various $p$ for $H$ $\parallel$ $ab$  (a) and $H$ $\parallel$ $c$ (b).
(c)-(e) $T$ dependencies of  $^{77}$Se NMR 1/$T_1T$ (left axes) at various $p$ with  $H$ $\parallel$ $ab$  (magenta circles) and $H$ $\parallel$ $c$ (gray circles). The arrows indicate $T_{\rm c}$ under 7.4089 T \cite{SM}. The green curve for each panel (right axes) shows the $T$ dependence of $C{K_ {\rm spin,c}^2}$, where $K_{\rm spin,c}$ is the spin part of $K_{\rm c}$ estimated by subtracting the $T$-independent $K_0$ and $C$ is a scaling factor. For all $p$, 0.175$\%$ and 9.5 are used for $K_0$ and $C$,  respectively. The insets in (c)-(e) show the $T$ dependence of the ratio $r \equiv (1/T_1T)_{ab}/(1/T_1T)_c$. The two horizontal lines represent the expected values for stripe-type ($r$ = 1.5) and Néel-type ($r$ = 0.5) AFMSFs, respectively.}

\label{fig:Fig2}
\end{figure*}

  Single crystals of  FeSe$_{0.47}$Te$_{0.53}$ ($T_{\rm c}$ = 14.7 K at ambient $p$ under $H$ = 0) were prepared using the chemical vapor transport technique following Ref. \cite{Ishida2022}, and details are given in the Supplemental Materials (SM) \cite{SM}.
  $^{125}$Te  ($I$ = 1/2, $\gamma_{\rm N}/2\pi$ = 13.454 MHz/T) NMR was measured at ambient $p$, while $^{77}$Se ($I$ = 1/2, $\gamma_{\rm N}/2\pi$ = 8.118 MHz/T) NMR was measured up to 1.35 GPa \cite{Pressure}.
     NMR spectra were obtained by sweeping frequency at $\mu_0H$ = 7.4089 T or sweeping $H$ at constant frequencies. 
     1/$T_{\rm 1}$ was measured with a saturation recovery method \cite{T1}.
    $T_{\rm c}$ was determined by ac susceptibility measurements \cite{SM}.
 Magnetic susceptibility was measured at $\mu_0H$ = 7 T  using a SQUID magnetometer (Quantum Design, MPMS).


 Figure~\ref{fig:Fig1}(b) shows the typical $^{77}$Se and $^{125}$Te NMR spectra of FeSe$_{0.47}$Te$_{0.53}$ in the normal state at $T$ = 15 K and 70 K for $H$ parallel to the  $ab$ plane ($H$ $\parallel$ $ab$) and $H$ parallel to the $c$ axis ($H$ $\parallel$ $c$) ($T_{\rm c}$ = 13.7 K (12.5 K) under $H$ $\parallel$ $ab$ ($H$ $\parallel$ $c$) of 7.4089 T \cite{SM}). 
  All spectra observed are a little  asymmetric, which could be due to a slight distribution of the hyperfine coupling constants $A$ of the Se and Te sites.  
   The full width at half maximum (FWHM) of the $^{77}$Se NMR spectra decrease almost linearly from  85  and 120 Oe at  100 K to 53  and 94 Oe at 15 K for  $H$ $\parallel$ $ab$ and $H$ $\parallel$ $c$, respectively. 
   A similar $T$ dependence of the FWHM is observed in the  $^{125}$Te NMR spectra where the FWHM decreases from 183  and  242 Oe at  100 K to 100  and 137 Oe at 15 K for  $H$ $\parallel$ $ab$ and $H$ $\parallel$ $c$, respectively. 
     The FWHM of $^{125}$Te NMR spectra is 1.5-2.1 as large as that of $^{77}$Se NMR spectra, which is mainly due to the different $A$, as shown below. 
    It is also noted that the FWHM of the $^{77}$Se NMR spectra in FeSe$_{0.47}$Te$_{0.53}$ are 3$\sim$5 times that in FeSe$_{0.0.71}$Te$_{0.29}$, due to the higher inhomogeneity in FeSe$_{1-x}$Te$_{x}$ as mentioned above (See the comparison in SM).

      The $T$ dependencies of $^{77}$Se and $^{125}$Te Knight shifts ($^{77}K$ and $^{125}K$) under  $H$ $\parallel$ $ab$  ($K_{ab}$) and $H$ $\parallel$ $c$  ($K_c$) are shown in Fig.~\ref{fig:Fig1}(c). 
  Both $K_{ab}$ and $K_c$ for $^{77}$Se and $^{125}$Te nuclei decrease with decreasing $T$ and exhibit  further reduction below $T_{\rm c}$, suggesting a spin-singlet SC state.     
  This $T$ dependence is qualitatively consistent with previous studies on similar compounds of $x$ = 0.58 \cite{Arčon2010}, 0.5 \cite{Shimizu2009}, 0.67 \cite{Michioka2010}, and 0.6 \cite{Hara2011} in FeSe$_{1-x}$Te$_x$. 
   The $T$ dependencies of $K_{ab}$ and $K_c$ of $^{77}$Se and $^{125}$Te are also consistent with those of the magnetic susceptibilities $\chi(T)$ shown in Figs.~\ref{fig:Fig1}(d) and \ref{fig:Fig1}(e), except for the low $T$ where the upturns were observed.  
   The upturns in $\chi(T)$ at low $T$, therefore,  can be not intrinsic and arise from a small amount of paramagnetic impurities. 
   The impurity contributions were assumed to be of Curie form and were subtracted from the observed  $\chi(T)$ so as to match the $T$ dependencies of $^{77}K$ with the corrected  $\chi(T)$ shown by open circles [see Figs.~\ref{fig:Fig1}(d) and \ref{fig:Fig1}(e)]. 
    In fact, the linear relation between the Knight shifts $K$ and the corrected $\chi$ can be seen in the  insets of Figs.~\ref{fig:Fig1}(d) and \ref{fig:Fig1}(e) where  $K_{ab}$ and $K_c$ for $^{77}$Se and $^{125}$Te are plotted against the corresponding corrected $\chi_{ab}$ and $\chi_c$, respectively. 
   
    As  $K$ is the sum of the $T$-dependent spin part $K_{\rm spin}$ and a $T$-independent orbital part $K_0$ (that is,  $K(T) = K_0 + K_{\rm spin}$) and  $K_{\rm spin}$ is proportional to the spin susceptibility $\chi_{\rm spin}$ through the hyperfine coupling constants $A$, the slope provides an estimate of $A$.  
     From those slopes, the values of $A$  for $^{125}$Te ($^{77}$Se) are estimated to be  $^{125}A_{ab}$ = 7.0(7) T/$\mu_{\rm B}$ ($^{77}A_{ab}$ = 4.3(4) T/$\mu_{\rm B}$) and   $^{125}A_c$ = 7.2(7) T/$\mu_{\rm B}$ ($^{77}A_c$ = 4.1(4) T/$\mu_{\rm B}$) for $H$ $\parallel$ $ab$ and $H$ $\parallel$ $c$, respectively. 
These $A$ values are greater than the  $^{125}A_{ab}$ = 3.63 T/$\mu_{\rm B}$ and $^{125}A_c$ = 4.88 T/$\mu_{\rm B}$ ($^{77}A_c$ = 2.85 T/$\mu_{\rm B}$)  in Fe$_{1.04}$Se$_{0.33}$Te$_{0.67}$ \cite{Michioka2010} but are comparable to $^{77}A_{ab}$ = 3.585 T/$\mu_{\rm B}$ and $^{77}A_c$ = 4.37 T/$\mu_{\rm B}$ in FeSe \cite{Wiecki2018}.
.
  
  To investigate the dynamical properties of the compound, we measured  1/$T_1T$ for both nuclei under the two $H$ directions whose results are shown in Fig.~\ref{fig:Fig1}(f). 
  Although the 1/$T_1T$ values of  $^{77}$Se and $^{125}$Te are different due to the difference in the $A$ values and the nuclear gyromagnetic ratio,  1/$T_1T$ for both nuclei shows a similar $T$ dependence where 1/$T_1T$  decreases as $T$ decreases and becomes nearly constant below $\sim$ 50 K within our experimental uncertainty.
    Below $T_{\rm c}$ (shown by arrows in  Fig.~\ref{fig:Fig1}(f)), 1/$T_1T$'s for both nuclei decrease exponentially due to the opening of SC gap,  which may suggest a fully gapped SC state. 
     In fact, as shown in the inset of  Fig.~\ref{fig:Fig1}(f), an exponential decrease of 1/$T_1$ can be seen in the semi-log plot of  $1/T_1$ vs. 1/$T$ from which the gap magnitude is estimated to be $\Delta$/$k_{\rm B}T_{\rm c}$ = 2.4(3)  within our experimental $T$ range measured.
    A similar exponential behavior of 1/$T_1$ in the SC state has been reported in  FeSe$_{0.42}$Te$_{0.58}$  with a slightly larger gap magnitude of $\Delta$/$k_{\rm B}T_{\rm c}$ = 3.0  \cite{Arčon2010}. 
     The STM measurements also show a full-gap SC in optimally Te substituted FeSe \cite{Hanaguri2010}.
     However, the SC-gap structure is still controversial. 
     A nodal gap structure has been suggested from  $^{125}$Te NMR measurements in  Fe$_{1.04}$Se$_{0.33}$Te$_{0.67}$ where 1/$T_1$ $\propto$ $T^5$ just below $T_{\rm c}$ has been reported  \cite{Michioka2010}. 
    In addition, a SC state with multiple full gaps has been observed in FeSe$_{0.45}$Te$_{0.55}$ by the later STM study \cite{Sarkar2017}.   
    Further studies are required to clarify the SC gap structure of FeSe$_{1-x}$Te$_{x}$ around $x$ = 0.5-0.6.

    Now we discuss the magnetic fluctuations in the normal state, focusing on how  it changes with $p$ because $T_{\rm c}$ largely depends on $p$ \cite{SM,Mukasa2021}.
     For this purpose, we carried out $^{77}$Se NMR measurements under $p$ of  0.55 and 1.35 GPa which leads to an increase in $T_{\rm c}$ from 14.7 K at ambient $p$ to 17.6 K at 0.55 GPa, and to 19.8 K at 1.35 GPa at $H$ = 0 (see details in the SM \cite{SM}).   
      Since $^{125}$Te-$1/T_1T$ and   $^{77}$Se-$1/T_1T$ show a similar $T$ dependence as described above, here we discuss the SFs using only  $^{77}$Se-$1/T_1T$ data. 
    Generally, 1/$T_1T$ is related to the ${\bf q}$-sum dynamical magnetic susceptibility, while the NMR shift $K$ measures the uniform magnetic susceptibility $\chi'({\bf q} = 0,\omega_{\rm N} = 0)$.
   Therefore, by comparing the $T$ dependence of 1/$T_1T$ and $K$, one can obtain insight into the $T$ evolution of $\sum_{\rm \bf q} \chi''({\bf q},\omega_{\rm N})$ with respect to that of $\chi'(0, 0)$. 
   To obtain the $T$ dependence of  NMR shift at different  pressures, we measured   $^{77}$Se NMR spectra of  FeSe$_{0.47}$Te$_{0.53}$ (see typical observed spectra  shown in the insets in Fig.~\ref{fig:Fig2}) under $p$. 
  The obtained $T$ and $p$ dependencies of the $^{77}K_{ab}$ and $^{77}K_{c}$ are shown in  Figs.~\ref{fig:Fig2}(a) and \ref{fig:Fig2}(b), respectively, showing that $^{77}K_{ab}$ and $^{77}K_{c}$ are nearly $p$ independent. Given the nearly $p$-independent values of $1/T_1T$ at high-temperature region as described below,  together with the nearly $p$-independent behavior of Knight shift, the hyperfine coupling constants are considered to be  nearly $p$-independent up to at least 1.35 GPa, as in the case of the isostructural  FeSe$_{1-x}$S$_{x}$ \cite{Rana2020,Rana2022,Rana2023,Kuwayama2019,Kuwayama2020,Kuwayama2021}.
  It is interesting to point out that, since $K$ is proportional to the density of states at the Fermi energy $N({\rm E_F})$,  the results indicate that $N({\rm E_F})$ is nearly independent of $p$, although $T_{\rm c}$ changes significantly. 
   This is in contrast to conventional BCS superconductors, in which $N({\rm E_F})$ generally correlates with $T_{\rm c}$.
   These results suggests that the SC in FeSe$_{1-x}$Te$_x$ is unconventional rather than conventional 
  which is consistent with the results of the Raman spectroscopy measurements that reveal the strength of electron-phonon coupling is insufficient to produce $T_{\rm c}$ $\sim$ 14 K in  FeSe$_{0.4}$Te$_{0.6}$\cite{Wu2020}.

   Here we compare the $T$ dependencies of  1/$T_1T$  with those of $K$.   
      Figures~\ref{fig:Fig2}(c)--\ref{fig:Fig2}(e) display  the $T$ dependencies of 1/$T_1T$ for both $H$ $\parallel$ $ab$ and $H$ $\parallel$ $c$  at ambient pressure,  0.55 GPa, and 1.35 GPa, respectively.
    At ambient $p$ as described above, there is a qualitative similarity in the $T$ dependencies of 1/$T_1T$ and $K$ above $\sim $ 40 K.
    However,  below this $T$, one can see the different $T$ dependence between 1/$T_1T$ and $K$ where  1/$T_1T$ for both field directions becomes nearly constant, while $K_{ab}$ and $K_{c}$ keep decreasing as shown in Fig.~\ref{fig:Fig2}(a) and \ref{fig:Fig2}(b).
    The deviation at low $T$ can be attributed to the development of AFMSF with ${\bf q} \ne$ 0.      
    Interestingly, at $p$ = 0.55 GPa, the large enhancements in 1/$T_1T$ are observed at low $T$ below $\sim$ 60 K  although less $p$ effects were detected at high $T$.  
      A further enhancement in 1/$T_1T$  at low $T$ can be seen at $p$ = 1.35 GPa,  indicating a strong enhancement of the AFMSF under pressure \cite{T1enhancement}.   
   According to previous NMR studies performed on IBSCs and their related materials  \cite{Kitagawa2009,Kitagawa2010,Hirano2012,Furukawa2014,Pandey2013,Ding2016}, one can obtain the information on the nature of SFs from the ratio $r \equiv T_{1,c}/T_{1,ab}$.   In most IBSCs, $r$ has been found to be $\sim$1.5, corresponding to the stripe-type AFMSF with ${\bf q} = (\pi,0)$ or $(0,\pi)$.   On the other hand, $r$ = 0.5 is expected for ${\bf q} = (\pi,\pi)$  spin correlations.   Note the wave vectors are given in the single-iron Brillouin zone notation.     As shown in the insets in Figs.~\ref{fig:Fig2}(c)-(e), $r$ is in the range of 1 $\sim$ 1.5 for all the pressures, but never decreases to 0.5.     This suggests  that the AFMSFs are characterized with ${\bf q} = (\pi,0)$ or $(0,\pi)$.  This is consistent with the inelastic neutron scattering study which  revealed dominant stripe-type AFMSF below $\sim$ 100 K in FeSe$_{1-x}$Te$_{x}$ although bicollinear stripe-type AFMSF also contribute at higher $T$ above 100 K \cite{Xu2016}.

    To explore the relationship between  the magnitude of AFMSF and $T_{\rm c}$, we follow our previous papers for FeSe$_{1-x}$S$_{x}$ \cite{Rana2020,Rana2022,Rana2023} where the observed 1/$T_1T$ was decomposed into the  two contributions: $\bf q$-independent component $(1/T_1T)_0$ and AFM component  $(1/T_1T)_{\rm AFM}$.
    $(1/T_1T)_0$ is expected to be proportional to ${K_ {\rm spin}^2}$. 
    Therefore, by comparing the observed 1/$T_1T$ with ${K_ {\rm spin}^2}$,  $(1/T_1T)_{\rm AFM}$ can be estimated. 
    Since  (1/$T_1T)_{ab}$ is enhanced more than  (1/$T_1T)_{c}$, which is expected for the case of strip-type AFM spin correlations, we compare the (1/$T_1T)_{ab}$ with $K_c$.
    The green curve for each panel (right axes) in Figs.~\ref{fig:Fig2}(c)-(e)  shows the $T$ dependence of $C{K_ {\rm spin, \it c}^2}$, where $C$ is a scaling factor. 
    At ambient $p$, a good scaling between  1/$T_1T$ and $C{K_ {\rm spin,c}^2}$ at high $T$ can be seen in Fig.~\ref{fig:Fig2}(c), using $K_0$ = 0.175$\%$ and $C$ = 9.5. 
    Those values seem to be reasonable as similar $C$ values (7.75-8.5) and the same $K_0$  have been reported in FeSe$_{1-x}$S$_{x}$ \cite{Rana2020,Rana2023}. 
    Since there is no clear change in $K_c$ with pressure, we used the same values of $K_0$ and $C$  for $p$ =  0.55 and 1.35 GPa.  
     Thus, the differences between  (1/$T_1T)_{ab}$ (magenta circles) and the green curves are attributed to the contributions of AFM spin fluctuations, $(1/T_1T)_{\rm AFM}$.

  \begin{figure}[tb]
  \includegraphics[width=\columnwidth]{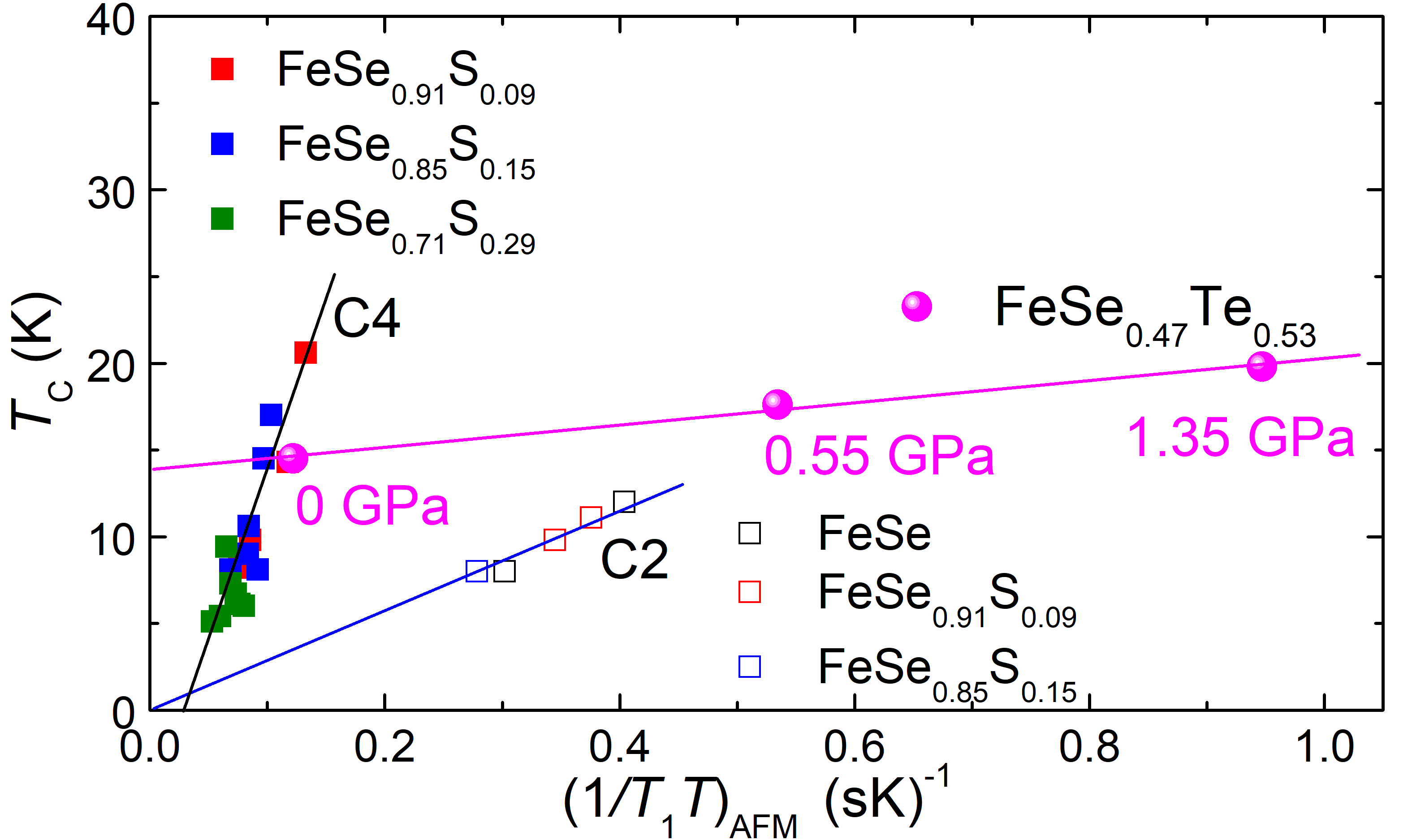}
	 \caption{Plot of $T_{\rm c}$ at zero field vs (1/$T_1T)_{\rm AFM}$. (1/$T_1T)_{\rm AFM}$ value is taken at $T$ around 20 $\sim$ 30 K in FeSe$_{0.47}$Te$_{0.53}$ and $T$ = 15 K in FeSe$_{1-x}$S$_{x}$ from Refs. \cite{Rana2020,Rana2023}.
Black and blue lines show linear fits for the $C_4$ (closed symbols) and $C_2$ (open symbols) phases of FeSe$_{1-x}$S$_{x}$, respectively. 
 }
	\label{fig:Fig3}
\end{figure}
 
Figure~\ref{fig:Fig3}  presents the relationship between   $T_{\rm c}$ under zero $H$ and  the magnitude of AFMSF under different $p$. 
   Here the magnitudes of AFMSF are represented by the values of  $(1/T_1T)_{\rm AFM}$ just above $T_{\rm c}$.  
    For comparison, data from previous studies for FeSe$_{1-x}$S$_{x}$ were also plotted in Fig.~\ref{fig:Fig3} \cite{Rana2020,Rana2022,Rana2023}.
    In FeSe$_{1-x}$S$_{x}$, it was revealed that $T_{\rm c}$ is proportional to AFMSFs in the $C_4$ (closed symbols) and $C_2$ (open symbols) phases as shown by black and blue lines, respectively \cite{Rana2020,  Rana2023}. 
    It was also pointed out that the slope for the $C_4$ phase is steeper than that for the  $C_2$ state, indicating that the AFMSFs without nematicity enhance $T_{\rm c}$ more than those with nematicity \cite{Rana2020, Rana2023}. 
     The origin of different slopes has been discussed in terms of the different numbers of hotspots on Fermi surfaces with and without nematicity \cite{Rana2023,Wang2017}.
   Interestingly, both the $C_2$ and  $C_4$ lines for FeSe$_{1-x}$S$_{x}$ start from around the origin  in the plot, suggesting the SC in FeSe$_{1-x}$S$_{x}$ is mainly induced by AFMSFs.

    In the case of FeSe$_{0.47}$Te$_{0.53}$, both the AFMSFs and $T_{\rm c}$ enhance with $p$ and a linear relationship between them is found as shown by the magenta line in Fig.~\ref{fig:Fig3}.
However, the slope of this relationship is quite different from those observed in FeSe$_{1-x}$S$_{x}$. 
    The small slope cannot be simply explained by the number of hotspots as the Fermi surface topology of the very close compound FeSe$_{0.45}$Te$_{0.55}$ \cite{Li2024} has been reported to be very similar to the case of the $C_4$ phase in FeSe$_{1-x}$S$_{x}$ \cite{Rana2023}.
     Thus the smaller slope  indicates that AFMSFs do not contribute significantly  to the SC in FeSe$_{0.47}$Te$_{0.53}$, which is quite different from the case of FeSe$_{1-x}$S$_{x}$. 
Furthermore,  the intercept of the magenta line in the plot is significantly away from the origin, indicating a high residual $T_c$ value even as AFMSFs approach zero.
  This behavior implies that the mechanism of SC in FeSe$_{0.47}$Te$_{0.53}$ is not primarily driven by AFMSFs but likely has another origin which would be naturally attributed to nematic fluctuations. 
   This interpretation aligns with the observation of a pure nematic QCP and the enhancement of the SC pairing strength near this nematic QCP \cite{Ishida2022,Mukasa2023}.   

 Recently the importance of nematicity or nematic fluctuations has been actually pointed out in many IBSCs \cite{Toyoda1,Dai2015,Yang2015,Kissikov2017,Curro2022,Toyoda2,Malinowski2020,Dioguardi2016,Eckberg2020,Wiecki2017,WangP2017,Bohmer2018,Hosoi2016,Mukasa2021}.
For example, in LiFe$_{1-x}$Co$_{x}$As, the strongest spin fluctuations were observed in the sample around $x$ $\sim$ 0.12, far from the maximum $T_{\rm c}$  position \cite{Dai2015}. 
In addition, in the highly-doped LaFeAsO$_{1-x}$F$_{x}$, AFMSFs are found to be weak but $T_{\rm c}$ is high \cite{Yang2015}. 
These results imply a different SC mechanism other than AFMSF.
The close relation between enhanced pairing and nematic fluctuations was observed in Ba$_{1-x}$Sr$_{x}$Ni$_2$As$_2$ \cite{Eckberg2020}. 
Furthermore, the suppression of $T_{\rm c}$ by anisotropic strain near a nematic QCP in Ba(Fe$_{1-x}$Co$_{x}$)$_2$As$_2$ \cite{Malinowski2020} also provides strong evidence of the SC driven by nematic fluctuations. 
  Nevertheless, the relationship between the magnitude of nematic fluctuations and $T_c$ seems not yet well established. 
  Therefore, our findings strongly call for further detailed investigation of the relationship in not only FeSe$_{0.47}$Te$_{0.53}$ under pressure but also other systems. 
  Such studies utilizing nematicity sensitive measurements including elastoresistivity and Nernst measurements \cite{Wuttke2022} will be essential to elucidate the contributions of nematic fluctuations to SC in unconventional SCs.

   In summary,  high pressure NMR measurements of $^{77}$Se and $^{125}$Te have been performed on FeSe$_{0.47}$Te$_{0.53}$ , which is close to a nematic QCP.    
   By comparing the correlations between the magnitude of AFM spin fluctuations and $T_{\rm c}$ in  FeSe$_{0.47}$Te$_{0.53}$ and FeSe$_{1-x}$S$_{x}$, nematic fluctuations were suggested to play an important role for the SC in FeSe$_{0.47}$Te$_{0.53}$.    
   The different contributions of AFMSF to SC in S and Te substituted FeSe were pointed out to be the origin of the different behavior of $T_{\rm c}$ around nematic QCP for each system  \cite{Mukasa2023}.     
   Our results under pressure highlight the FeSe$_{1-x}$Te$_{x}$ system as a very suitable platform to study the role of nematic fluctuations on SC, stimulating further investigations of the system under pressure, and also provide insights into the mechanism of unconventional superconductivity. 
Systematic NMR measurements at ambient and under pressure on FeSe$_{1-x}$Te$_{x}$ with different Te contents below and above the nematic QCP are in progress.

We would like to acknowledge C. L. Mueller for assistance in the preparation of the crystals. The research was supported by the U.S. Department of Energy (DOE), Office of Basic Energy Sciences, Division of Materials Sciences and Engineering. Ames National Laboratory is operated for the U.S. DOE by Iowa State University under Contract No.~DE-AC02-07CH11358.


\begin{thebibliography}{99}

\bibitem{Kamihara2008} Y. Kamihara, T. Watanabe, M. Hirano, and H. Hosono, Iron-Based Layered Superconductor La[O$_{1-x}$F$_{x}$]FeAs ($x$ = 0.05--0.12) with $T_{\rm c}$ = 26 K, J. Am. Chem. Soc. {\bf 130}, 3296 (2008).
\bibitem{Johnston2010} D. C. Johnston, The puzzle of high temperature superconductivity in layered iron pnictides and chalcogenides, Adv.  Phys. {\bf 59}, 803 (2010).
\bibitem{Canfield2010} P. C. Canfield and S. L. Bud'ko, FeAs-Based Superconductivity: A Case Study of the Effects of Transition Metal Doping on BaFe$_2$As$_2$, Annu. Rev. Condens. Matter Phys. {\bf 1}, 27 (2010).
\bibitem{Stewart2011} G. R. Stewart, Superconductivity in iron compounds, Rev. Mod. Phys. {\bf 83}, 1589 (2011).
\bibitem{Cui2017} J. Cui, Q.-P. Ding, W. R. Meier, A. E. Böhmer, T. Kong, V. Borisov, Y. Lee, S. L. Bud'ko, R. Valentí, P. C. Canfield, and Y. Furukawa, Magnetic fluctuations and superconducting properties of CaKFe$_4$As$_4$ studied by  $^{75}$As NMR, Phys. Rev. B \textbf{96}, 104512 (2017).
\bibitem{Meier2018} W. R. Meier, Q.-P. Ding, A. Kreyssig, S. L. Bud’ko, A. Sapkota, K. Kothapalli, V. Borisov, R. Valentí, C. D. Batista, P. P. Orth, R. M. Fernandes, A. I. Goldman, Y. Furukawa, A. E. Böhmer, and P. C. Canfield, Hedgehog spin-vortex crystal stabilized in a hole-doped iron-based superconductor, npj Quantum Mater.  {\bf 3}, 5 (2018).
\bibitem{Wilde2023} J.M. Wilde, A. Sapkota, Q.P. Ding, M. Xu, W. Tian, S.L. Bud’ko, Y. Furukawa, A. Kreyssig, P.C. Canfield, Antiferromagnetic order and its interplay with superconductivity in CaK(Fe$_{1-x}$Mn$_{x}$)$_4$As$_4$, J. Phys.: Condens. Matter {\bf 35}, 395801 (2023).
\bibitem{Xu2023} M. Xu, J. Schmidt, M. A. Tanatar, R. Prozorov, S. L. Bud'ko, and P. C. Canfield, Superconductivity and magnetic and transport properties of single-crystalline CaK(Fe$_{1-x}$Cr$_{x}$)$_4$As$_4$, Phys. Rev. B {\bf 107}, 134511 (2023).
\bibitem{Ding2018} Q.-P. Ding, W. R. Meier, J. Cui, M. Xu, A. E. Böhmer, S. L. Bud’ko, P. C. Canfield, and Y. Furukawa, Hedgehog Spin-Vortex Crystal Antiferromagnetic Quantum Criticality in CaK(Fe$_{1-x}$Ni$_{x}$)$_4$As$_4$ Revealed by NMR, Phys. Rev. Lett. \textbf{121}, 137204 (2018).
\bibitem{Bohmer2018} A. E. B\"{o}hmer and A. Kreisel, Nematicity, magnetism and superconductivity in FeSe, J. Phys. Condens. Matter {\bf 30}, 023001 (2018).
\bibitem{Hosoi2016}S. Hosoi, K. Matsuura, K. Ishida, H. Wang, Y. Mizukami, T. Watashige, S. Kasahara, Y. Matsuda, and T. Shibauchi, Nematic quantum critical point without magnetism in FeSe$_{1-x}$S$_x$ superconductors, Proc. Natl. Acad. Sci. USA {\bf 113}, 8139 (2016).
\bibitem{Coldea2019} A. I. Coldea, S. F. Blake, S. Kasahara, A. A. Haghighirad, M. D. Watson, W. Knafo, E. S. Choi, A. McCollam, P. Reiss, T. Yamashita, M. Bruma, S. Speller, Y. Matsuda, T. Wolf, T. Shibauchi, and A. J. Schofield, Evolution of the Fermi surface of the nematic superconductors FeSe$_{1-x}$S$_x$, Npj Quant. Mater. {\bf 4}, 2 (2019).
\bibitem{Mukasa2021} K. Mukasa, K. Matsuura, M. Qiu, M. Saito, Y. Sugimura, K. Ishida, M. Otani, Y. Onishi, Y. Mizukami, K. Hashimoto, J. Gouchi, R. Kumai, Y. Uwatoko, and T. Shibauchi, High-pressure phase diagrams of FeS$_{1-x}$Te$_{x}$: Correlation between suppressed nematicity and enhanced superconductivity, Nat. Commun. \textbf{12}, 381 (2021).
\bibitem{Ishida2022} K. Ishida, Y. Onishi, M. Tsujii, K. Mukasa, M. Qiu, M. Saito, Y. Sugimura, K. Matsuura, Y. Mizukami, K.
Hashimoto, and T. Shibauchi, Pure Nematic Quantum Critical Point Accompanied by a Superconducting Dome, Proc. Natl. Acad. Sci. U.S.A. \textbf{119}, e2110501119 (2022).
\bibitem{Hsu2008} F.-C. Hsu, J.-Y. Luo, K.-W. Yeh, T.-K. Chen, T.-W. Huang, P. M. Wu, Y.-C. Lee, Y.-L. Huang, Y.-Y. Chu, D.-C. Yan, and M.-K. Wu, Superconductivity in the PbO-type structure $\alpha$-FeSe, Proc. Natl. Acad. Sci. U.S.A. {\bf 105}, 14262 (2008). 
\bibitem{McQueen2009} T. M. McQueen, A. J. Williams, P. W. Stephens, J. Tao, Y. Zhu, V. Ksenofontov, F. Casper, C. Felser, and R. J. Cava, Tetragonal-to-Orthorhombic Structural Phase Transition at 90 K in the Superconductor Fe$_{1.01}$Se, Phys. Rev. Lett. {\bf 103}, 057002 (2009).
\bibitem{Abdel2015}M. Abdel-Hafiez, Y.-Y. Zhang, Z.-Y. Cao, C.-G. Duan, G. Karapetrov, V. M. Pudalov, V. A. Vlasenko, A. V. Sadakov, D. A. Knyazev, T. A. Romanova, D. A. Chareev, O. S. Volkova, A. N. Vasiliev, and X.-J. Chen, Superconducting properties of sulfur-doped iron selenide, Phys. Rev. B {\bf 91}, 165109 (2015).
\bibitem{Watson2015} M. D. Watson, T. K. Kim, A. A. Haghighirad, S. F. Blake, N. R. Davies, M. Hoesch, T. Wolf, and A. I. Coldea, Suppression of orbital ordering by chemical pressure in FeSe$_{1-x}$S$_x$, Phys. Rev. B {\bf  92}, 121108(R) (2015).
\bibitem{Reiss2017} P. Reiss, M. D. Watson, T. K. Kim, A. A. Haghighirad, D. N. Woodruff, M. Bruma, S. J. Clarke, and A. I. Coldea, Suppression of electronic correlations by chemical pressure from FeSe to FeS, Phys. Rev. B {\bf 96}, 121103(R) (2017).
\bibitem{Lai2015} X. Lai, H. Zhang, Y. Wang, X. Wang, X. Zhang, J. Lin, and F. Huang, Observation of Superconductivity in Tetragonal FeS, J. Am. Chem. Soc., {\bf 137} 10148 (2015).


\bibitem{Böhmer2015} A. E. Böhmer, T. Arai, F. Hardy, T. Hattori, T. Iye, T. Wolf, H. v. Löhneysen, K. Ishida, and C. Meingast, Origin of the Tetragonal-to-Orthorhombic Phase Transition in FeSe: A Combined Thermodynamic and NMR Study of Nematicity, Phys. Rev. Lett. \textbf{114}, 027001 (2015).
\bibitem{Baek2015} S-H. Baek, D.V. Efremov, J.M. Ok, J.S. Kim, J. van den Brink, and B. Büchner, Orbital-driven nematicity in FeSe, Nat. Mater. \textbf{14}, 210  (2015).
\bibitem{Imai2009} T. Imai, K. Ahilan, F. L. Ning, T. M. McQueen, and R. J. Cava, Why Does Undoped FeSe Become a High-$T_c$ Superconductor under Pressure? Phys. Rev. Lett. \textbf{102}, 177005 (2009).
\bibitem{Wiecki2017} P. Wiecki, M. Nandi, A. E. Böhmer, S. L. Bud'ko, P. C. Canfield, and Y. Furukawa, NMR evidence for static local nematicity and its cooperative interplay with low-energy magnetic fluctuations in FeSe under pressure, Phys. Rev. B   \textbf{96}, 180502(R) (2017).
\bibitem{Wiecki2018} P. Wiecki, K. Rana, A. E. Böhmer, Y. Lee, S. L. Bud'ko, P. C. Canfield, and Y. Furukawa, Persistent correlation between superconductivity and antiferromagnetic fluctuations near a nematic quantum critical point in FeSe$_{1-x}$S$_{x}$, Phys. Rev. B \textbf{98}, 020507(R) (2018).
\bibitem{Rana2020} K. Rana, L. Xiang, P. Wiecki, R. A. Ribeiro, G. G. Lesseux, A. E. Böhmer, S. L. Bud’ko, P. C. Canfield, and Y. Furukawa, impact of nematicity on the relationship between antiferromagnetic fluctuations and superconductivity in FeSe$_{0.91}$S$_{0.09}$ under pressure,  Phys. Rev. B \textbf{101}, 180503(R) (2020).
\bibitem{Rana2022} K. Rana, and Y. Furukawa, Relationship between Nematicity, Antiferromagnetic Fluctuations, and Superconductivity in  FeSe$_{1-x}$S$_{x}$ revealed by NMR,  Front. Phys. \textbf{10}, 849284 (2022).
\bibitem{Rana2023} K. Rana, D. V. Ambika, S. L. Bud'ko, A. E. Böhmer, P. C. Canfield, and Y. Furukawa, Interrelationships between nematicity, antiferromagnetic spin fluctuations, and superconductivity: Role of hotspots in FeSe$_{1-x}$S$_{x}$ revealed by high pressure $^{77}$Se NMR study,  Phys. Rev. B \textbf{107}, 134507 (2023).
\bibitem{Kuwayama2020}T. Kuwayama, K. Matsuura, Y. Mizukami, S. Kasahara, Y. Matsuda, T. Shibauchi, Y. Uwatoko, and N. Fujiwara, NMR study under pressure on the iron-based superconductor  FeSe$_{1-x}$S$_{x}$ ($x$ = 0.12 and 0.23): Relationship between nematicity and AF fluctuations, Mod. Phys. Lett.  B {\bf 34}, 2040048 (2020).
\bibitem{Kuwayama2021}T. Kuwayama, K. Matsuura, J. Gouchi, Y. Yamakawa, Y. Mizukami, S. Kasahara, Y. Matsuda, T. Shibauchi, H. Kontani, Y. Uwatoko and N. Fujiwara, Pressure-induced reconstitution of Fermi surfaces and spin fluctuations in S-substituted FeSe, Sci. Rep. {\bf 11}, 17265 (2021).
\bibitem{Kuwayama2019}T. Kuwayama, K. Matsuura, Y. Mizukmami, S. Kasahara, Y. Matsuda, T. Shibauchi, Y. Uwatoko and N. Fujiwara, $^{77}$Se-NMR Study under Pressure on 12$\%$-S Doped FeSe, J. Phys. Soc. Jpn. {\bf 88}, 033703 (2019).
\bibitem{Yu2023} Z. Yu, K. Nakamura, K. Inomata, X. Shen, T. Mikuri, K. Matsuura, Y. Mizukami, S. Kasahara, Y. Matsuda, T. Shibauchi $et al$., Spin fluctuations from Bogoliubov Fermi surfaces in the superconducting state of S-substituted FeSe, Commun. Phys. \textbf{6}, 175 (2023).
\bibitem{Mukasa2023} K. Mukasa, K. Ishida, S. Imajo, M. Qiu, M. Saito, K. Matsuura, Y. Sugimura, S. Liu, Y. Uezono, T. Otsuka, M. Čulo, S. Kasahara, Y. Matsuda, N. E. Hussey, T. Watanabe, K. Kindo, and T. Shibauchi, Enhanced Superconducting Pairing Strength near a Pure Nematic Quantum Critical Point,  Phys. Rev. X \textbf{13}, 011032 (2023).
\bibitem{Jiang2023} Q. Jiang, Y. Shi, M. H. Christensen, J. Sanchez, B. Huang, Z. Lin, Z. Liu, P. Malinowski, X. Xu, R. M. Fernandes, and J.-H. Chu, Nematic fluctuations in an orbital selective superconductor Fe$_{1+y}$Te$_{1-x}$Se$_x$. Commun. Phys. \textbf{6}, 39 (2023).
\bibitem{Otsuka2019} T. Otsuka, S. Hagisawa, Y. Koshika, S. Adachi, T. Usui, N. Sasaki, S. Sasaki, S. Yamaguchi, Y. Nakanishi, M. Yoshizawa, S. Kimura, and T. Watanabe, Incoherent-coherent crossover and the pseudogap in Te-annealed superconducting Fe$_{1+y}$Te$_{1-x}$Se$_{x}$ revealed by magnetotransport measurements, Phys. Rev. B \textbf{99}, 184505 (2019).
\bibitem{Hou2023} Q. Hou, L. Sun, Y. Sun, and Z. Shi, Review of Single Crystal Synthesis of 11 Iron-Based Superconductors, Materials {\bf 16}, 4895 (2023).
\bibitem{Yin2015} J. X. Yin, Z. Wu, and J. H. Wang $et$ $al$., Observation of a robust zero-energy bound state in iron-based superconductor Fe(Te,Se), Nat. Phys. \textbf{11}, 543 (2015).
\bibitem{Wang2018} D. Wang, L. Kong, and P. Fan $et$ $al$., Evidence for Majorana bound states in an iron-based superconductor, Science \textbf{362}, 333 (2018).
\bibitem{Machida2019} T. Machida, Y. Sun, S. Pyon, S. Takeda, Y. Kohsaka, T. Hanaguri, T. Sasagawa, and T. Tamegai, Zero-energy vortex bound state in the superconducting topological surface state of Fe(Se,Te), Nat. Mater. \textbf{18}, 811 (2019).
\bibitem{Zhang2018} P. Zhang, K. Yaji, and T. Hashimoto $et$ $al$., Observation of topological superconductivity on the surface of an iron-based superconductor, Science \textbf{360}, 182 (2018).
\bibitem{Li2024} Y.-F. Li, S.-D. Chen, M. García-Díez, M. I. Iraola, H. Pfau, Y.-L. Zhu, Z.-Q. Mao, T. Chen, M. Yi, P.-C. Dai, J. A. Sobota, M. Hashimoto, M. G. Vergniory, D.-H. Lu, and Z.-X. Shen, Orbital Ingredients and Persistent Dirac Surface State for the Topological Band Structure in FeTe$_{0.55}$Se$_{0.45}$, Phys. Rev. X \textbf{14}, 021043 (2024).
 \bibitem{SM} See Supplemental material for the details of sample preparation, EDS measurements, the results of the ac susceptibility measurements, the comparison of the $^{77}$Se NMR spectra to those of FeSe$_{0.71}$S$_{0.29}$, NMR pulse conditions in the SC state, and the typical relaxation curves for $^{77}$Se $T_1$ measurements.
\bibitem{Pressure} The pressure was applied with a NiCrAl/CuBe piston-cylinder cell using Daphne 7373 as the pressure transmitting medium.  Pressure calibration was accomplished by $^{63}$Cu nuclear quadruple resonance in Cu$_2$O \cite{Reyes1992,Fukazawa2007} at 77 K.
\bibitem{Reyes1992}  A. P. Reyes, E. T. Ahrens, R. H. Heffner, P. C. Hammel, and J. D. Thompson, Cuprous oxide manometer for high‐pressure magnetic resonance experiments, Rev. Sci. Instrum. \textbf{63}, 3120 (1992). 
\bibitem{Fukazawa2007} H. Fukazawa, N. Yamatoji, Y. Kohori, C. Terakura, N. Takeshita, Y. Tokura, and H. Takagi, Manometer extension for high pressure measurement: Nuclear quadrupole resonance study of Cu$_2$O with a modified Bridgman anvil cell up to 10 GPa, Rev. Sci. Instrum. \textbf{78}, 015106 (2007).

\bibitem{T1}  $1/T_1$ at each $T$ is determined by fitting the nuclear magnetization $M$ versus time $t$ dependence after saturation using the exponential function $1-$$M$($t$)/$M({\infty})$ = exp[${-(t/T_1)^\beta]}$.
   Here $M$($t$) and $M({\infty})$ are the nuclear magnetization at time $t$ after saturation and the equilibrium nuclear magnetization at time $t$ $\rightarrow$ $\infty$, respectively. 
   $\beta$ was found to be unity in the normal state, and starts decreasing in the superconducting state, reaching $\sim$ 0.7 at low temperatures.
   Different from the previous reports in which the recovery curves in $T_1$ measurement in FeTe$_{1-x}$Se$_{x}$ show non-single-exponential behavior \cite{Arčon2010,Shimizu2009,Michioka2010,Hara2011}, the recovery curves in this study show the single-exponential behavior in the normal state. See details in the Supplemental Materials  \cite{SM}.

\bibitem{Arčon2010} D. Arčon, P. Jeglič, A. Zorko, A. Potočnik, A. Y. Ganin, Y. Takabayashi, M. J. Rosseinsky, and K. Prassides,
Coexistence of localized and itinerant electronic states in the multiband iron-based superconductor FeSe$_{0.42}$Te$_{0.58}$, Phys. Rev. B \textbf{82}, 140508(R) (2012).
\bibitem{Shimizu2009} Y. Shimizu, T. Yamada, T. Takami, S. Niitaka, H. Takagi, and M. Itoh, Pressure-Induced Antiferromagnetic Fluctuations in the Pnictide Superconductor FeSe$_{0.5}$Te$_{0.5}$: $^{125}$Te NMR Study, J. Phys. Soc. Jpn. \textbf{78}, 123709 (2009).
\bibitem{Michioka2010} C. Michioka, H. Ohta, M. Matsui, J. Yang, K. Yoshimura, and M. Fang, Macroscopic physical properties and spin dynamics in the layered superconductor Fe$_{1+\delta}$Te$_{1-x}$Se$_{x}$, Phys. Rev. B \textbf{82}, 064506 (2010).
\bibitem{Hara2011} Y. Hara, H. Kotegawa, H. Nohara, H. Tou, Y. Mizuguchi, and Y. Takano, Se/Te-NMR study of Fe(Te$_{1-x}$Se$_{x}$), J. Phys. Soc. Jpn. \textbf{80}, SA119 (2011).
\bibitem{Hanaguri2010} T. Hanaguri, S. Niitaka, K. Kuroki, and H. Takagi, Unconventional $s$-wave superconductivity in Fe(Se,Te), Science \textbf{328}, 474 (2010).
\bibitem{Sarkar2017} S. Sarkar, J. Van Dyke, P. O. Sprau, F. Massee, U. Welp, W.-K. Kwok, J. C. S. Davis, and D. K. Morr, Orbital superconductivity, defects, and pinned nematic fluctuations in the doped iron chalcogenide FeSe$_{0.45}$Te$_{0.55}$, Phys. Rev. B \textbf{96}, 060504(R) (2017).
\bibitem{Wu2020} S.-F.Wu, A. Almoalem, I. Feldman, A. Lee, A. Kanigel, and G. Blumberg, Superconductivity and Phonon Self-Energy Effects in Fe$_{1+y}$Te$_{0.6}$Se$_{0.4}$, Phys. Rev. Res.  \textbf{2}, 013373 (2020).
\bibitem{T1enhancement} A similar enhancement of 1/$T_1$ of $^{125}$Te under pressure has been reported in a powder form of FeSe$_{0.5}$Te$_{0.5}$ in Ref. \cite{Shimizu2009}.
\bibitem{Kitagawa2009} K. Kitagawa, N. Katayama, K. Ohgushi, and M. Takigawa, Antiferromagnetism of SrFe$_2$As$_2$ Studied by Single-Crystal $^{75}$As-NMR,  J. Phys. Soc. Jpn. \textbf{78}, 063706 (2009).
\bibitem{Kitagawa2010} S. Kitagawa, Y. Nakai, T. Iye, K. Ishida, Y. Kamihara, M. Hirano, and H. Hosono, Stripe antiferromagnetic correlations in LaFeAsO$_{1-x}$F$_{x}$ probed by $^{75}$As NMR, Phys. Rev. B \textbf{81}, 212502 (2010).
\bibitem{Hirano2012} M. Hirano, Y. Yamada, T. Saito, R. Nagashima, T. Konishi, T. Toriyama, Y. Ohta, H. Fukazawa, Y. Kohori, Y. Furukawa, K. Kihou, C.-H. Lee, A. Iyo, and H. Eisaki,  Potential Antiferromagnetic Fluctuations in Hole-Doped Iron-Pnictide Superconductor Ba$_{1-x}$K$_{x}$Fe$_2$As$_2$ Studied by $^{75}$As Nuclear Magnetic Resonance Measurement, J. Phys. Soc. Jpn. \textbf{81}, 054704 (2012).
\bibitem{Furukawa2014} Y. Furukawa, B. Roy, S. Ran, S. L. Bud’ko, and P. C. Canfield,  Suppression of electron correlations in the collapsed tetragonal phase of CaFe$_2$As$_2$ under ambient pressure demonstrated by $^{75}$As NMR/NQR measurements, Phys. Rev. B \textbf{89}, 121109(R) (2014).
\bibitem{Pandey2013} A. Pandey, D. G. Quirinale, W. Jayasekara, A. Sapkota, M. G. Kim, R. S. Dhaka, Y. Lee, T. W. Heitmann, P. W. Stephens, V. Ogloblichev, A. Kreyssig, R. J. McQueeney, A. I. Goldman, A. Kaminski, B. N. Harmon, Y. Furukawa, and D. C. Johnston, Crystallographic, electronic, thermal, and magnetic properties of single-crystal SrCo$_2$As$_2$,  Phys. Rev. B \textbf{88}, 014526 (2013).
\bibitem{Ding2016} Q.-P. Ding, P. Wiecki, V. K. Anand, N. S. Sangeetha, Y. Lee, D. C. Johnston, and Y. Furukawa, Volovik effect and Fermi-liquid behavior in the $s$-wave superconductor CaPd$_2$As$_2$: $^{75}$As NMR-NQR measurements, Phys. Rev. B \textbf{93}, 140502(R) (2016).
\bibitem{Xu2016} Z. Xu, J. A. Schneeloch, J. Wen, E. S. Božin, G. E. Granroth, B. L. Winn, M. Feygenson, R. J. Birgeneau, G. Gu, I. A. Zaliznyak, J. M. Tranquada, and G. Xu, Thermal evolution of antiferromagnetic correlations and tetrahedral bond angles in superconducting FeTe$_{1-x}$Se$_{x}$,  Phys. Rev. B \textbf{93}, 104517 (2016).

\bibitem{Wang2017} X. Wang, Y. Schattner, E. Berg, and R. M. Fernandes, Superconductivity mediated by quantum critical antiferromagnetic fluctuations: The rise and fall of hot spots, Phys. Rev. B \textbf{95}, 174520 (2017).
\bibitem{Toyoda1} M. Toyoda, Y. Kobayashi, and M. Itoh, Nematic fluctuations in iron arsenides NaFeAs and LiFeAs probed by  $^{75}$As NMR, Phys. Rev. B {\bf 97}, 094515 (2018).
\bibitem{Dai2015} Y. M. Dai, H. Miao, L. Y. Xing, X. C. Wang, P. S. Wang, H. Xiao, T. Qian, P. Richard, X. G. Qiu, W. Yu, C. Q. Jin, Z. Wang, P. D.  Johnson, C. C. Homes, and H. Ding, Spin-Fluctuation-Induced Non-Fermi-Liquid Behavior with Suppressed Superconductivity in LiFe$_{1-x}$Co$_{x}$As, Phys. Rev. X \textbf{5}, 031035 (2015).
\bibitem{Yang2015} J. Yang, R. Zhou, L.-L. Wei, H.-X. Yang, J.-Q. Li, Z.-X. Zhao, and G.-Q. Zheng, New Superconductivity Dome in LaFeAsO$_{1-x}$F$_x$ Accompanied by Structural Transition, Chin. Phys. Lett. {\bf 32}, 107401 (2015).
\bibitem{Kissikov2017} T. Kissikov, R. Sarkar, M. Lawson, B. T. Bush, E. I. Timmons, M. A. Tanatar, R. Prozorov, S. L. Bud'ko, P. C. Canfield, R. M. Fernandes, W. F. Goh, W. E. Pickett, and N. J. Curro, NMR study of nematic spin fluctuations in a detwinned single crystal of underdoped Ba(Fe$_{1-x}$Co$_{x}$)$_2$As$_2$, Phys. Rev. B {\bf 96}, 241108(R) (2017).
\bibitem{Curro2022} N. J. Curro, T. Kissikov, M. A. Tanatar, R. Prozorov, S. L. Bud’ko, and P. C. Canfield,  Nematicity and Glassy Behavior Probed by Nuclear Magnetic Resonance in Iron-Based Superconductors, Front. Phys. {\bf 10}, 877628 (2022).
\bibitem{Toyoda2} M. Toyoda, A. Ichikawa, Y. Kobayashi, M. Sato, and M. Itoh, In-plane anisotropy of the electric field gradient in Ba(Fe$_{1-x}$Co$_{x}$)$_2$As$_2$ observed by  $^{75}$As NMR, Phys. Rev. B {\bf 97}, 174507 (2018).
\bibitem{Malinowski2020} P. Malinowski, Q. Jiang, J. J. Sanchez, J. Mutch, Z. Liu, P. Went, J. Liu, P. J. Ryan, J. W. Kim, and J. H. Chu, Suppression of superconductivity by anisotropic strain near a nematic quantum critical point, Nat. Phys. \textbf{16}, 1189 (2020).
\bibitem{Dioguardi2016} A. P. Dioguardi, T. Kissikov, C. H. Lin, K. R. Shirer, M. N. Lawson, H.-J. Grafe, J.-H. Chu, I. R. Fisher, R. M. Fernandes, and N. J. Curro, NMR Evidence for Inhomogeneous Nematic Fluctuations in BaFe$_2$(As$_{1-x}$P$_{x}$)$_2$, Phys. Rev. Lett.  {\bf 116}, 107202 (2016).
\bibitem{Eckberg2020} C. Eckberg, D. J. Campbell, T. Metz, J. Collini, H. Hodovanets, T. Drye, P. Zavalij, M. H. Christensen, R. M. Fernandes, S. Lee,
P.  Abbamonte, J. W. Lynn, and J. Paglione,  Sixfold enhancement of superconductivity in a tunable electronic nematic system, Nat. Phys. \textbf{16}, 346 (2020).
\bibitem{WangP2017} P. S. Wang, P. Zhou, S. S. Sun, Y. Cui, T. R. Li, H. Lei, Z. Wang, and W. Yu, Robust short-range-ordered nematicity in FeSe evidenced by high-pressure NMR, Phys. Rev. B {\bf 96}, 094528 (2017).
\bibitem{Wuttke2022} C. Wuttke, F. Caglieris, S. Sykora, F. Steckel, X. Hong, S. Ran, S. Khim, R. Kappenberger, S. L. Bud’ko, P. C. Canfield, S. Wurmehl, S. Aswartham, B. Büechner, and C. Hess, Ubiquitous enhancement of nematic fluctuations across the phase diagram of iron based superconductors probed by the Nernst effect,  npj Quantum Mater. \textbf{7}, 82 (2022).



\end{thebibliography}
\end{document}